# How Does the National New Area Impact the Local Economy? —An empirical analysis from Zhoushan

Yi Zheng

Business School, Shanghai Jian Qiao University, Shanghai 201306，China.
Email: yzheng@shou.edu.cn

**Abstract:** In order to empirically study the policy impact of a National New Area on the local economy, this paper evaluates the effect of the Zhoushan Archipelago New Area on local GDP growth rate and economic efficiency. Firstly, collecting input and output data of 20 prefectural-level cities in Jiangsu, Zhejiang and Anhui provinces from 1995 to 2015, the economic efficiency of these 20 cities is estimated by data envelopment analysis. Then selecting some cities from the above cities except Zhoushan, we construct some counterfactuals of Zhoushan by a panel data approach. The difference between the actual and counterfactual values for GDP growth rate and economic efficiency of Zhoushan is compared to conclude the treatment effect of National New Area. The research shows that in the first four years, the policy of New Area promoted Zhoushan's economic quality by raising its efficiency, but relatively reduced its economic quantity by negatively affected local GDP growth rate. Then the policy influence on Zhoushan's GDP growth rate and economic efficiency is gradually disappeared after four years. Combining some other research on our study, we find that the policy effect on GDP growth rate is related to the level of economic growth in these areas. The policy of New Area has less influence on GDP growth if they are approved in the developed region rather than in undeveloped region. We think that it is better to set up a New Area in a relatively undeveloped zone.

## 1 Introduction

As the result of the global flow of production factors and changing of production organization, regional economic environment gradually become an appropriate unit to take part in global competition for promoting their countries' economic growth. In China evolution of development policies to improve local economic environment can be divided into three stages: the exploration stage led by Special Economic Zones, the expansion stage dominated by Economic Development Zones and the optimization stage featuring National New Areas (NNAs) and National Comprehensive Reform Pilot Areas (Qi et al., 2017). China's NNA (Guojia Xinqu) approved by the State Council should be as a functional urban areas, which undertake the strategic tasks of national development and reform (Chao et al., 2015). It is also becoming a policy tool for China's opening and transformation, and an important way to integrate resources and cultivate functions. Supported by government policies, the NNAs have gathered a large number of innovative elements of institution and economy. It wants to promote the transformation of industrial structure to modern industrial system by quality improvement, and to enter one of middle and high level of global production value chain by efficiency change (Zhang et al., 2018). But in practice, do the NNAs enhance the quality and efficiency of economic development and make it become the internal power of local economic agglomeration and industrial structure upgrading? This needs to be verified by actual cases. The NNA is a typical place-based industrial policy. It will also provide new evidence for the current industrial policy debate.

Since the starting of the 21st century, China's opening up and economic development has been facing severe challenges. At the same time, with the process of urbanization crossing the inflection point, the dividend of urbanization is disappearing rapidly and the sustainable development of urbanization is also facing challenges. To solve these problems, Chinese government is setting up more NNAs to deepen the opening up and promote regional economic development through their "first-act and first-try". Till June 2019, the number of NNAs has been increased to 19 in China and their importance to the economy has also on the rise. Nevertheless, different NNAs seems to have different effects on their local economy. For example, Pudong New Area in Shanghai has greatly promoted regional economic growth in the past three decades. (Alder et al., 2016）On the other hand, some research showed that the concentration of productive resources by policy advantages of NNAs may result in the lack of resources outside the host areas. This will cause a negative impact on the regional economy. (Lin, et al., 2018) So, whether the overall impact of a NNA on local economic growth was positive or negative need for more empirical research.. At the same time, timely evaluation on the influence is conducive to promoting the NNA's local economy and to provide recommendations for NNAs' layout and adjustment.

NNA is an economic policy with China's identifiable characteristics (Chao et al., 2015). Although the practice of development zones has expanded worldwide, there are few studies on NNAs in foreign academic circles. In China, literature for NNAs was scarce before 2016. With the increase of NNAs in recent years, the related research has increased significantly since 2016. Among them, there are only a few studies on the impact of NNAs on their local economic growth. Some researchers (Chao et al., 2018, Wang et al., 2019) analyzed NNAs at district and county level by diverse methods. They found that the studied NNAs had promoted local economic development, but their impact on the regional economy varies with their geographical location and establishment time. Fan et al. (2017) thought that some of NNAs, such as Shanghai Pudong New Area, Tianjin Binhai New Area and Zhejiang Zhoushan Archipelago New Area (ZANA), have formed regional growth poles and their effective radiation radius have reached 800 km, he (Fan al.et,. 2018) also founded that some NNAs in south China have some negative impact on their local economic growth rate.

Above researches mainly focused on the quantity, such as GDP, when evaluating the impact of regional economic development. But at the national level, the only meaning of "competitiveness" is the productivity of the country, and the productivity of the country is mainly reflected in the economic efficiency of large cities and node cities on the transportation network (Gao, 2008). So analysis of these cities' economic efficiency can reflect the quality of indigenous economic development and the national competitiveness. However, in the majority of existing literature which evaluated the economic impact of NNAs, the quality of economic development hasn't been taken into consideration. Only a few studies have been of the view that it. But there are some deficiencies in them. For instance, Wang et al. (2018) used DEA method to calculate the economic efficiency of the enterprises in Tianjin in the year of 2004 and 2008, one of them was before and another is after the foundation of Tianjin Binhai New Area in 2006. Then they thought that Tianjin Binhai New Area had an effect on improving the local enterprises' efficiency. As the cross-sectional data used for analysis are only from one year before and after the New Area, their conclusion is lacking in persuasion. Using the data between the year of 2000 and 2005, Lin et al. (2018) studied the impact of some national economic development zones on total factor productivity of enterprises. Because the needed data of this research is only obtained from the year of 2000 to 2005, its conclusion has reflected the situation of the early national economic development zones. Now the internal and external environment for NNAs has changed with them. So whether the conclusion of this paper is suitable for NNAs after 2005 needs further verification.

Based on the above analysis, on the one hand, different NNAs seem to have different impacts on local economic quantity. The reliability of the conclusion needs more cases to support. On the other hand, research on impact of NNAs on local economic quality is still relatively few and imperfect. So evaluating more the impact of NNAs on the regional economy in quantity and quality can give more cases to verify the above conclusion and make up for the weaknesses of the related research on efficiency. It is a meaningful work for research.

In June 2011, Chinese State Council formally allowed to set up ZANA in Zhejiang Province. Zhoushan has become the first National Archipelago New Area in China. Zhoushan covers 22,000 square kilometers of sea area and 1,371 square kilometers of land area in east China. It is a prefecture-level city (Dijishi) in China with archipelagic structure, including 1,390 islands, 270km coastlines and 1.1 million populations. It has a small land area and large sea area. Its focus of economic development is in the marine economy. Therefore, founding of ZANA is an important policy attempt for promoting economy of coastal areas in China. Analyzing its actual effects will be helpful not only for the evaluating of NNAs, but also for the development of China's marine economy. So in this paper we study whether ZANA has promoted local economy in quantity and quality.

In Fig.1 we find that the GDP growth rate of Zhoushan did not increase significantly after 2010, when ZANA was founded (on the right side of the vertical line in Fig.1). Considering there are many factors affecting economic development, the chart can't directly prove ZANA has no impact on the local economy. To study the impact, the other affecting factors should be stripped, despite the fact that it is difficult sometimes. In addition, to study the impact on the economy, we should consider not only its quantity represented by the GDP growth rate, but also its quality represented by economic efficiency.

Chinese process of economic reforms launched in 1978, and gradually extended until current days. Because a variety of new different policies and institutions were introduced simultaneously, even today it is difficult to pinpoint their exact effect. This paper try to analyze ZANA by the panel data approach to remove the impact of other factors and calculate efficiency by the data envelopment analysis (DEA) to consider economic quality. This will contribute to a better understanding of the place-based policies. It also provides further evidence in the debate about the effect of these policies. So the main contributions of this paper are that: Firstly, using the panel data approach, a new way of evaluating the impact of NNAs is provided. To a certain extent, it solves the endogenous problem in the empirical study of NNAs. Secondly, the impact of ZANA on the local economy is studied from the aspects of quantity and quality. It enriches empirical cases on the study of policy effects of NNAs. Thirdly, Research on the impact of NNAs on economic efficiency has always been a weak link in the existing research. By combining the panel data approach with DEA method, this paper helps to remedy this shortcoming.

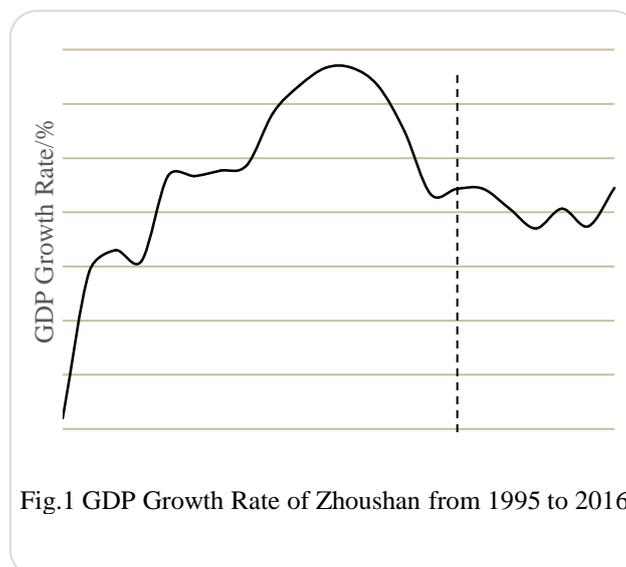

Fig.1 GDP Growth Rate of Zhoushan from 1995 to 2016

The rest of the paper is organized as following. Section 2 presents the research method and the sample. In section 3, using the economic data from 1995 to 2015, the super efficiency of Zhoushan and other 19 selected prefectural-level cities are evaluated by DEA. Section 4 empirically studies the impact of ZANA on GDP growth rate and economic efficiency by the panel data approach. Section 5 concludes and discusses.

## 2  Method and data

### 2.1  Selection of method

For predictive models to provide reliable guidance in decision making processes,they are often required to be accurate and robust to distribution shifts (Zheng et al., 2022).To evaluate the impact on the regional economy, most practices usually construct a regression model by some selected economic indicators. The commonly used methods based on the regression are simultaneous equations model, VAR model and DSGE model. However, these methods have many strict assumptions and are difficult to satisfy. For example, the simultaneous equations model relies on the prior assumptions of exogenous and endogenous variables, while the VAR model has the limitation of variable dimension and it is difficult for explaining the economic mechanism. Sometimes, economists apply quasi-experimental methods to macro-policy evaluation. The method regards a project or policy as an experiment and tries to find a natural control group, which is under well-designed experimental conditions to its experimental group. Quasi-experimental methods include instrumental variables (IV), differences in differences (DID), and regression discontinuity (RD). There are some restrictions on their use. For example, DID method must satisfy the assumption of randomness, which means that all uncontrollable factors should change randomly to unbiased influence the samples in the experimental group and the control group. Sometimes it's hard.

To improve these methods, Hsiao et al. (2010) proposed a panel data approach for measuring policy effects. This approach considers the economic operation of each body in the cross-section to be driven by some common factors from the same economic system. Although the driven degree may be different for each body, there is certain linkage or correlation between these cross-sections. So the counterfactual values of an individual without the policy affecting can be predicted after it has been affected actually by the policy. The panel data approach relaxes the assumption of randomness of DID method. It can also overcome the difficulties of ambiguous causality, complex theoretical modeling, missing variables and insufficient time series data in macro-policy evaluation. To some extent, it can reduce the interference of variable selection and estimation methods on the robustness of empirical results.

Although the panel data approach has not been put forward for a long time, it has been widely and successfully applied in the evaluation of policy effects. Using this approach, Hsiao et al. (2010) assessed the impact of Hong Kong's political and economic integration with the mainland after its return；Ching et al (2011) assessed the impact of China's accession to the WTO; Zhao et al. (2017) estimated the future economic effects if the China-Pakistan Economic Corridor was connected; Tan et al. (2015), Yin et al. (2017) and Wang et al. (2017) respectively evaluated the effect of Shanghai Pilot Free Trade Zone on local economic growth. Ying et al. (2018) estimated the economic effect of the four Pilot Free Trade Zones in Shanghai, Tianjin, Fujian and Guangdong too. The above study has verified the applicability of the panel data approach in evaluation of macro-policy performance. So we choose the approach developed by Hsiao et al. (2012) to construct counterfactual economy of Zhoushan in ZANA for studying the impact of ZANA on the local economy. In addition, in order to analyze the impact of ZANA on economic efficiency, it is necessary to measure economic efficiency of these selected cities firstly. For the purpose, data envelopment analysis (DEA), which has been widely used in efficiency analysis, is used in this paper.

### 2.2  Determine of research scope

To study economy change by DEA and the panel data analysis, we select per capita real GDP as an output indicator, the number of employees in urban units at the end of the year, the area of land in administrative area, public financial expenditure and investment in fixed assets as four input indicators. The data are sourced from "China City Statistical Yearbooks".

In June 2011, ZANA has been formally set up. In December 2015, the State Council approved again to establish China (Zhejiang) Bulk Commodity Exchange Center in Zhoushan, and in august 2016, Zhoushan Free Trade Experimental Zone has been established. These two events have impacted Zhoushan's economy with ZANA. In order to avoid their interference, only five-year data from 2011 to 2015 is collected to reflect the economy in ZANA. On the other hand, the panel data approach requires the period before the policy to be as long as possible than the ones in the policy. So 16 years of data from 1995 to 2010 are collected to reflect the economy before ZANA. Now collected data not only meet the requirements of the approach, but also reduce the interference from other economic factors besides ZANA.

Assuming that an individual in an experimental group is affected by a policy, the individuals in a control group are not affected by the policy. For estimating a counterfactual value of the individual in the experimental group, which is assumed not to be affected although it is actually affected by the policy, the panel data approach requires that at the same time cross-section, the individuals in the experimental group and control group are driven by some common factors (such as population, capital, technology, etc.) in their development. Although these factors have different effects on different individual, it makes these individuals have some kind of correlation in their economy. Thus, the regression of the individuals in the control group can be used to predict the counterfactual values of the individual in the experimental group (Hsiao et al., 2012).

China's administrative system has five hierarchical levels of government: (1) central; (2) provincial; (3) prefecture; (4) county; and (5) township. Zhoushan is a prefecture-level city, an administrative division ranking below a province and above a county in China's administrative structure. Generally, a prefecture-level city comprises a core urban area and a surrounding periphery that may include rural areas, other smaller cities, towns and villages. The geographical and administrative position of the prefecture-level cities in Jiangsu, Zhejiang and Anhui is close to Zhoushan and therefore these cities have a similar economic environment as Zhoushan. According to the panel data approach, there is a correlation between these cities and Zhoushan. So Zhoushan which has been affected by the policy of ZANA, is selected as an experimental group in this paper, and some of prefecture-level cities in Jiangsu, Zhejiang and Anhui provinces might be selected to compose a control group for explaining the bulk of Zhoushan economy.

As for the selection principles of these cities, first of all, they need complete economic data from 1995 to 2015. Some cities are excluded because their data is broken for changing in administrative divisions during the research period. For example, Chaohu City was abolished and divided into Hefei, Wuhu and Ma'anshan in Anhui Province in 2011. In order to avoid statistical error, these four prefecture-level cities are excluded. Secondly, according to the panel data approach, the economy of the selected cities should not be affected by the policy of ZANA, so Shanghai, Ningbo and other cities are also excluded because of their close relationship with Zhoushan. Finally, 19 cities in Jiangsu, Zhejiang and Anhui including Wuxi, Changzhou, Suzhou, Nantong, Lianyungang, Yancheng, Yangzhou, Zhenjiang, Wenzhou, Huzhou, Jinhua, Quzhou, Bengbu, Huainan, Huaibei, Tongling, Anqing, Huangshan and Chuzhou are left to choose. Now we need to calculate the economic efficiency of the 19 cities and Zhoushan for further analysis.

# 3 Evaluation of economic efficiency

This section calculates the economic efficiency of the 20 cities selected above. The results not only can help these cities promoting their resources to be rational organized and allocated, but also are used in the next section for analyzing the quality of economic development in Zhoushan.

## 3.1 Super efficiency DEA model

The DEA method has been widely used in efficiency evaluation because it is suitable for multi-input and multi-output data. It was proposed by Charnes et al.（1978）for evaluating the relative effectiveness between decision-making units (DMUs). In DEA model, if some DMUs are effective at the same time, i.e., they all have maximum efficiency value θ=1, they can't be further distinguished in efficiency. To make up for this deficiency, Anderson et al.（1993） created the super-efficiency DEA model. Its difference from traditional DEA model is that a DMU is excluded from the possible production set when its efficiencyis evaluated. Therefore in the super-efficiency DEA model, for an ineffective DMU, the frontier of the possible production set is consistent with the traditional DEA model, and its final efficiency value is the same as the value measured by the traditional DEA model. But for a effective DMU, whose efficiency value measured by the traditional DEA modelis 1, its production frontier of the super-efficiency DEA model may move backward comparing to the traditional DEA model, so its efficiency value measured by the super-efficiency DEA model may be larger than 1. Therefore, the super-efficiency DEA model can further identify the differences between the effective DMUs,whoseefficiency values areall equal to onein the traditional DEA models. It has overcome the shortcoming of traditional DEA model and now has been more widely used than a traditional DEA model.

For DMU q (q = 1,2,...,), the super-efficiency DEA model can be expressed as:

$$\min \theta_q$$

s.t. $\begin{cases} \sum_{j=1, j \neq q}^{n} x_{i,j} \lambda_j + s_i^- = \theta x_q, i = 1,2,\ldots,m \\ \sum_{j=1, j \neq q}^{n} y_{k,j} \lambda_j - s_k^+ = y_q, k = 1,2,\ldots,r \\ \lambda_j \geq 0, j = 1,2,\ldots,q-1, q+1,\ldots,n \\ s_i^- \geq 0, s_k^+ \geq 0 \end{cases}$ （1）

In this formula, θ represents a DMU's efficiency, x and y are input and output variables respectively, and λ represents the combination proportion for a DMU's effective value. Here ∑λ>1，∑λ=1and ∑λ<1 respectively indicate the increasing returns to scale, the constant returns to scale and the decreasing returns to scale, n is the number of DMUs in the system, m and r are the numbers of input and output variables respectively. $s_i^-$ and $s_k^+$ is relaxation variables, they indicate input excess and output deficit respectively. When θ is less than 1, the DMU is not effective and needs to be improved as in a traditional DEA model. When the values of θ for several DMUs are greater than or equal to 1, these DMUs are all effectiveand not be distinguished in a traditional DEA model, but they can be further ranked by their value θ in the super efficiency DEA model.

## 3.2 Evaluation of super efficiency

To calculate the economic efficiency of the selected prefecture-level cities, the number of employees in urban units at the end of the year, the area of land in administrative areas, public financial expenditure and fixed assets investment are chosen as four input indexes, and the GDP is taken as an output index according to the importance and availability of indicators. Such as, although the input of technology and knowledge also have impact on regional economic efficiency, their influence is significantly smaller than the four selected inputs, and their total amount is difficult to measure in macro-analysis, they generally won't be used as input indicators in the study of efficiency. So strictly speaking, the economic efficiency calculated by the selected input-output indexes in this paper means the economic efficiency of the main input.

By above design, more than 2,200 relevant data is collected from China Urban Statistics Yearbook for Zhoushan, Wuxi, Changzhou, Suzhou, Nantong, Lianyungang, Yancheng, Yangzhou, Zhenjiang, Wenzhou, Huzhou, Jinhua, Quzhou, Bengbu, Huainan, Huaibei, Tongling, Anqing, Huangshan and Chuzhou between the year of 1996 and 2016. Usingthe super-efficiency model shown in Formula 1, the economic efficiency of the selected 20 cities in 21 years is calculated by EMS software. The results are listing inTable 1.

Tab.1 Economic efficiency of 20 cities in Jiangsu, Zhejiang and Anhui from 1995 to 2015

| year | Zhoushan | Wuxi | Changzhou | Suzhou | Nantong | Lianyungang | Yancheng | Yangzhou | Zhenjiang | Wenzhou |
|---|---|---|---|---|---|---|---|---|---|---|
| 1995 | 0.70 | 1.54 | 0.95 | 1.27 | 0.85 | 0.56 | 0.96 | 0.95 | 1.04 | 1.08 |
| 1996 | 0.58 | 1.59 | 0.85 | 1.19 | 0.90 | 0.80 | 1.00 | 0.67 | 1.01 | 0.98 |
| 1997 | 0.59 | 1.55 | 0.90 | 1.11 | 0.82 | 0.68 | 0.91 | 0.78 | 0.89 | 0.79 |
| 1998 | 0.47 | 1.54 | 0.69 | 1.16 | 1.00 | 0.48 | 1.09 | 0.72 | 0.85 | 0.87 |
| 1999 | 0.32 | 0.86 | 1.18 | 0.55 | 0.93 | 1.13 | 1.21 | 1.11 | 1.40 | 0.45 |
| 2000 | 0.50 | 1.42 | 0.73 | 1.10 | 0.90 | 0.55 | 1.04 | 0.93 | 0.99 | 0.79 |
| 2001 | 0.54 | 1.41 | 0.93 | 1.03 | 1.04 | 0.58 | 1.45 | 0.79 | 1.06 | 0.92 |
| 2002 | 0.49 | 1.35 | 0.77 | 1.18 | 1.03 | 0.58 | 1.57 | 0.74 | 0.88 | 1.05 |
| 2003 | 0.70 | 1.20 | 0.87 | 0.92 | 0.93 | 0.71 | 0.97 | 0.94 | 1.10 | 0.98 |
| 2004 | 0.70 | 1.21 | 0.86 | 1.03 | 0.92 | 0.72 | 0.92 | 0.94 | 1.06 | 1.08 |
| 2005 | 0.76 | 1.28 | 0.85 | 1.00 | 0.98 | 0.60 | 0.85 | 0.92 | 1.03 | 1.27 |
| 2006 | 0.67 | 1.33 | 0.86 | 1.01 | 0.99 | 0.58 | 0.87 | 0.94 | 1.01 | 1.08 |
| 2007 | 0.62 | 1.33 | 0.85 | 1.03 | 0.96 | 0.52 | 0.79 | 0.92 | 1.04 | 1.11 |
| 2008 | 0.60 | 1.20 | 0.91 | 1.07 | 0.91 | 0.50 | 0.74 | 0.90 | 1.12 | 1.14 |
| 2009 | 0.56 | 1.14 | 0.94 | 1.21 | 0.91 | 0.51 | 0.71 | 0.91 | 1.12 | 1.10 |
| 2010 | 0.60 | 1.15 | 1.13 | 1.26 | 0.91 | 0.50 | 0.67 | 0.92 | 1.05 | 1.17 |
| 2011 | 0.64 | 1.18 | 0.89 | 1.14 | 0.84 | 0.53 | 0.69 | 0.86 | 0.98 | 0.83 |
| 2012 | 0.64 | 1.16 | 0.92 | 1.10 | 0.77 | 0.54 | 0.69 | 0.88 | 0.96 | 0.82 |
| 2013 | 0.80 | 1.31 | 0.97 | 1.10 | 0.77 | 0.64 | 0.74 | 0.90 | 1.00 | 0.81 |
| 2014 | 0.85 | 1.30 | 1.04 | 1.27 | 0.87 | 0.67 | 0.81 | 1.04 | 1.02 | 0.97 |
| 2015 | 0.46 | 1.28 | 1.05 | 1.35 | 0.79 | 0.62 | 0.69 | 0.85 | 0.92 | 0.78 |
| 平均 | 0.61 | 1.30 | 0.91 | 1.09 | 0.91 | 0.62 | 0.92 | 0.89 | 1.03 | 0.95 |

Continued Table 1

| year | Huzhou | Jinhua | Quzhou | Bengbu | Huainan | Huaibei | Tongling | Anqing | Huangshan | Chuzhou | range |
|---|---|---|---|---|---|---|---|---|---|---|---|
| 1995 | 0.84 | 1.53 | 0.82 | 0.81 | 0.30 | 0.45 | 0.34 | 0.65 | 0.44 | 0.97 | 1.23 |
| 1996 | 0.82 | 2.06 | 0.81 | 0.84 | 0.32 | 0.54 | 0.35 | 0.88 | 0.92 | 0.91 | 1.75 |
| 1997 | 0.87 | 2.16 | 0.86 | 1.13 | 0.59 | 0.78 | 0.65 | 0.98 | 0.97 | 1.50 | 1.57 |
| 1998 | 0.94 | 2.00 | 0.62 | 0.60 | 0.35 | 0.28 | 0.29 | 0.66 | 0.66 | 0.99 | 1.72 |
| 1999 | 0.54 | 0.78 | 0.35 | 0.37 | 0.33 | 0.17 | 0.24 | 0.37 | 0.41 | 0.58 | 1.24 |
| 2000 | 0.99 | 1.82 | 0.55 | 0.47 | 0.47 | 0.34 | 0.35 | 0.60 | 0.51 | 1.16 | 1.48 |
| 2001 | 0.94 | 1.37 | 0.59 | 0.57 | 0.75 | 0.55 | 0.45 | 0.68 | 0.53 | 1.23 | 0.99 |
| 2002 | 0.97 | 1.28 | 0.63 | 0.53 | 0.67 | 0.58 | 0.44 | 0.61 | 0.49 | 1.06 | 1.13 |
| 2003 | 0.93 | 0.88 | 0.63 | 0.75 | 1.08 | 0.83 | 0.77 | 0.77 | 0.64 | 1.47 | 0.83 |

| Year | | | | | | | | | | | |
|---|---|---|---|---|---|---|---|---|---|---|---|
| 2004 | 0.94 | 0.85 | 0.60 | 0.84 | 0.88 | 0.89 | 0.85 | 0.84 | 0.59 | 1.32 | 0.73 |
| 2005 | 0.88 | 0.89 | 0.63 | 0.86 | 0.72 | 0.84 | 0.91 | 0.94 | 0.59 | 0.98 | 0.69 |
| 2006 | 0.87 | 0.96 | 0.62 | 0.94 | 0.65 | 0.73 | 1.01 | 0.88 | 0.50 | 0.91 | 0.83 |
| 2007 | 0.89 | 1.07 | 0.63 | 0.75 | 0.59 | 0.68 | 0.91 | 0.67 | 0.44 | 0.67 | 0.89 |
| 2008 | 0.88 | 1.07 | 0.66 | 0.67 | 0.70 | 0.67 | 0.72 | 0.60 | 0.42 | 0.57 | 0.78 |
| 2009 | 0.83 | 1.01 | 0.59 | 0.56 | 0.66 | 0.58 | 0.58 | 0.49 | 0.40 | 0.47 | 0.81 |
| 2010 | 0.88 | 0.98 | 0.63 | 0.52 | 0.65 | 0.60 | 0.68 | 0.56 | 0.42 | 0.52 | 0.83 |
| 2011 | 0.87 | 1.14 | 0.71 | 0.50 | 0.59 | 0.55 | 0.71 | 0.58 | 0.48 | 0.56 | 0.70 |
| 2012 | 0.85 | 1.03 | 0.73 | 0.47 | 0.63 | 0.51 | 0.57 | 0.60 | 0.46 | 0.56 | 0.70 |
| 2013 | 0.95 | 0.97 | 0.82 | 0.62 | 0.47 | 0.57 | 0.69 | 0.65 | 0.64 | 0.77 | 0.84 |
| 2014 | 0.96 | 1.10 | 1.08 | 0.85 | 0.80 | 1.10 | 0.90 | 1.04 | 0.66 | 0.83 | 0.65 |
| 2015 | 0.76 | 0.77 | 0.76 | 0.68 | 0.56 | 0.53 | 0.76 | 0.59 | 0.62 | 0.73 | 0.89 |
| mean | 0.88 | 1.22 | 0.68 | 0.68 | 0.61 | 0.61 | 0.63 | 0.70 | 0.56 | 0.89 | 1.01 |

Note: Range is the difference between maximum and minimum efficiency in the same year.

In Table 1, the average efficiency of Wuxi, Suzhou, Zhenjiang and Jinhua is higher than 1. It means that they are more efficient in the 21 years than others. The first three of them are in Jiangsu province and the last is in Zhejiang province. We also find in the last column of the table 1 that the ranges between the highest and lowest efficiency of these cities are decreasing. For example, the range of efficiency between 1995 and 2000 is more than 1.2, while the range between 2003 and 2015 is no more than 0.89. Considering the relativity of efficiency measured by DEA model, this shows that the polarization of economic efficiency is reduced in the Yangtze River Delta in recent years. It may be indicated that the economic integration development in the Yangtze River Delta has promoted the rational allocation of resources. At the same time, it is noted that the economic efficiency of Zhoushan has been at a relatively low level in these years. Does it mean that the policy of ZANA has not promoted the economic efficiency of Zhoushan?

## 4 Analysis of the economic effect by ZANA

From Fig.1 and Table 1, we have seen the GDP growth rate and economic super-efficiency of Zhoushan not increasing significantly since 2010, when ZANA has been founded in Zhoushan. But this can't explain the actual role of the policy of ZANA, because an economic phenomenon is a combined result of many factors. In order to separate other factors besides ZANA, we construct a panel data approachas following.

### 4.1 Construction of the theoretical model

By the panel data approach (Hsiao et al., 2012), we assumed that the period of study is from t = 1 to t = T, the number of studied cities are N. An economic output of the city i (i=1,2,…,N) in year t is denoted as $y_{it}$. Zhoushan, as an experimental group, is expressed as the city 1 without loss of generality, namely i = 1 is for Zhoushan and $y_{1t}$ is its output. We will select from other N-1 cities to form a controlled group. The other N-1 cities are represented by i = 2, 3, …, N and their output in year t is expressed in $y_{2t}$. $y_{3t}$,…, $y_{Nt}$ respectively. $y_{it}^1$ and $y_{it}^0$ denote the economic output of the city i in year t with and without the policy intervention. Because ZANA was approved in June 2011, we assume $T_1$ is the year of 2010, so ZANA has taken effect in time $T_1+1$. Based on these assumptions, we have:

$$y_{1t} = y^0_{1t}, t = 1, \ldots, T_1; y_{1t} = y^1_{1t}, t = T_1 + 1, \ldots, T$$

As the total economic amount of Zhoushan is relatively small in Jiangsu, Zhejiang and Anhui, we think it has little impact on the economy of the other selected cities and ignores it in this paper. Therefore, whether or not the policy of ZANA was carried on in Zhoushan, the other selected cities almost had no impact. It denoted as following:

$$y_{it} = y^0_{it}, (i = 2, \ldots, N, t = 1, 2, \ldots, T)$$

Then, the effect of ZANA on Zhoushan's economy at time t is simply represented by

$$\Delta_{1t} = y^1_{1t} - y^0_{1t}, t = T_1 + 1, \ldots, T \qquad (2)$$

Where $\Delta_{1t}$ is the treatment effect of economic output by the policy of ZANA for Zhoushan. However, $y^1_{1t}$ and $y^0_{1t}$ can't be simultaneously observed. For the city i (i=1,2,…,N), the observed data are taken as the form of *(y_{it};d_{it})*,

$$y_{it} = d_{it} y^1_{it} + (1 - d_{it}) y^0_{it}, t = 1, \ldots, T \qquad (3)$$

Where *d_{it}=1* if the city i is in the policy treatment and *d_{it}=0* otherwise. Under the variable of policy intervention $d_{1t}$, the specific random components of regional economic growth in N-1 cities except Zhoushan are assumed conditionally independent, namely:

$$E(\varepsilon_{is} \mid d_{1t}) = 0, i = 2, \ldots, N, s \geq t$$

By formula (3), there wasn't any policy treatment to *y_{it}* for i =1, 2, 3, ..., N and t=1, . . . ,T_1. For t=T_1+1,…,T, the output of Zhousan, i.e. $y_{1t}$, is treated by the policy. All other output except Zhoushan, i.e. *y_{it}* (i=2, . . . ,N), don't be affected by the policy.

It has been documented empirically that there are common factors to explain the most macroeconomic data of the unit in the experimental group with selected units in the control group (Onatski, 2009). So for estimating the data $y^0_{1t}$ of Zhoushan in formula (2) after $T_1$, we assumed that the output of all selected cities can be decomposed into two components: the first is impacted by K common factors, $f_t$, which drove the output of all cities to change. K common factors may be national macro policies, international political and economic shocks, trade development, technological progress, etc. They may be unobserved or unknown. The second is the idiosyncratic components, $\alpha_i + \varepsilon_{it}$, where $\alpha_i$ represents the specific effect for the city i and $\varepsilon_{it}$ is the idiosyncratic error with $E(\varepsilon_{it})=0$ and uncorrelated with $\varepsilon_{jt}$ for j≠i. So the output of Zhoushan ($y_{1t}$) is captured by the common factors, the city's specific effect and the idiosyncratic error as follows.

$$y_{1t} = b_1 f_t + \alpha_1 + \varepsilon_{1t}, t = 1, 2, \ldots, T \qquad \ldots \ldots (4)$$

Among them, $f_t (K \times 1)$ is a K-dimensional column vector representing K common factors and changing with time t, $b_1 (K \times 1)$ denotes a K-dimensional row coefficient vector, which describes different influence degree of K common factors on Zhoushan. $\alpha_1$ represents the fixed effect of Zhoushan. $\varepsilon_{1t}$ is its random disturbance term and satisfied $E(\varepsilon_{1t}) = 0$.

The $f_t$ in formula (4) is not easy to observe in practice, but $(y_{1t}, ..., y_{Nt})$ are driven by these common factors $f_t$. So the panel data approach thinks there is a linear correlation between them. This makes Zhoushan's output $y_{1t}$ can be predicted by the other cities' outputs $(y_{2t}, ..., y_{Nt})$, i.e., the counterfactual output $y_{1t}^0$ of Zhoushan is predicted by the observed data $\tilde{y}_t^0 = (y_{2t}^0, ..., y_{Nt}^0)$ in lieu of $f_t$ in formula (4). So different regressions for Zhoushan are constructed with all different output combinations from the above selected 19 cities. Through comparing these regressions with some selection criterion, a combination of M cities is selected. M cities from a control group and their related optimal fitting are gotten as follows：

$$\hat{y}_{1t}^0 = \hat{\bar{\alpha}} + \hat{\alpha}_2 y_{2t}^0 + \ldots + \hat{\alpha}_M y_{Mt}^0, t = 1, \ldots T_1 \qquad \ldots\ldots\ldots(5)$$

Then substitute the observed output of the M cities into formula (5) between $T_1 + 1$ and T, the out-of-sample prediction of Zhoushan is has been gotten. The calculated results gave the counterfactual output of Zhoushan without interference of ZANA after the establishment of ZANA. Namely:

$$\hat{y}_{1t}^0 = \hat{\bar{\alpha}} + \hat{\alpha}_2 y_{2t}^0 + \ldots + \hat{\alpha}_M y_{Mt}^0, t = T_1 + 1, \ldots, T \ldots\ldots\ldots(6)$$

With the estimated $y_{1t}^0$ in formula (6), the affected effect $\Delta_{1t}$ of ZANA can be gotten by formula (2).

In above fitting process to get formula (5), the important problem is how to choose the best control group. Using more $y_{jt}$ may improve the within sample fit, but lead to inaccurate out-of-sample prediction. Hsiao et al. (2012) proved that the error of the affected effect estimated by formula (5) is better for some small samples. According to their research, the steps to find the best predictors in formula (5) are given as follows:

Step 1: For a fixed j (j = 1,2,...,N-1), we take any j individuals from the selected N-1 prefecture-level cities except Zhoushan. Combination of $C_{N-1}^j$ groups can be obtained. With each of these groups $(y_{2t}^0, y_{3t}^0, ..., y_{(j+1)t}^0)$, Zhoushan's output $(y_{1t}^0)$ is fitted by formula (4) in time t=1,2,...$T_1$. The best fitting group is selected by R-square and noted as $M^*(j)$. We repeated this process from 1 to N-1 for j and can selecte N-1 groups, they are noted as $M^*(j)$, for j = 1, 2,... N-1.

Step 2: In $M^*(j)$, for j = 1, 2,... N-1, a group $M^*$ will be chosen as a control group in terms of Akaike Information Criterion (AIC) (Akaike, 1973). By cities in $M^*$ the output of Zhoushan can be best fitted with formula (4) between time t=1 to $T_1$.

Step 3: Substituting the output of the cities in $M^*$ in time t=$T_1 + 1, \ldots, T$ into formula (4), the out-of-sample prediction of Zhoushan is gotten and presented by formula (5). It is the counterfactual value of Zhoushan without the interference of ZANA after ZANA has been approved. With the counterfactual value, the affected effect $\Delta_{1t}$ of ZANA could be estimated by formula (2).

## 4.2 Empirical analyses on the economy of Zhoushan
### 4.2.1 Empirical analysis on GDP growth rate

For analyzing the policy impact of ZANA on local economic quantity, we take GDP growth rate as an output index. According to the above three steps, firstly, before ZANA (1995-2010), for a fixed j (j = 1,2,..., 19), we take j cities from 19 selected cities to construct a group. There are $c_{19}^{j}$ different groups for a fixed j. the GDP growth rate of Zhoushan is fitted by each of them to get a best group with maximum correlation coefficient. Let's take j from 1 to 19, the 19 best groups are chosen. Secondly, we calculate AIC for each of the 19 selected groups and then among them choose a group with minimum AIC as a control group. The selected control group is composed in this paper by Changzhou ($X_1$), Nantong ($X_2$), Jinhua ($X_3$) and Tongling ($X_4$). Their fitting for Zhoushan is shown by formula (7). Fitting and choosing processes are completed by Pampe Package in R software and get the fitting formulaas follows.

$$Y=-3.797078+2.646954X_1-0.561032X_2-0.414065 X_3-0.302354 X_4 \quad (7)$$

Byformula (7), we get the fitted GDP growth rate of Zhoushan from 1995 to 2015. They are counterfactual values and are shown in Figure 2 with the actual values. In the pre-intervention period (the left side of the vertical line in Figure 2, i.e. before 2010), the fitted path produced by the selected cities in the control group closely adhered to the actual path of Zhoushan. So it's believable that the counterfactual GDP growth rate of Zhoushan can be deduced well by formula (7) after ZANA.

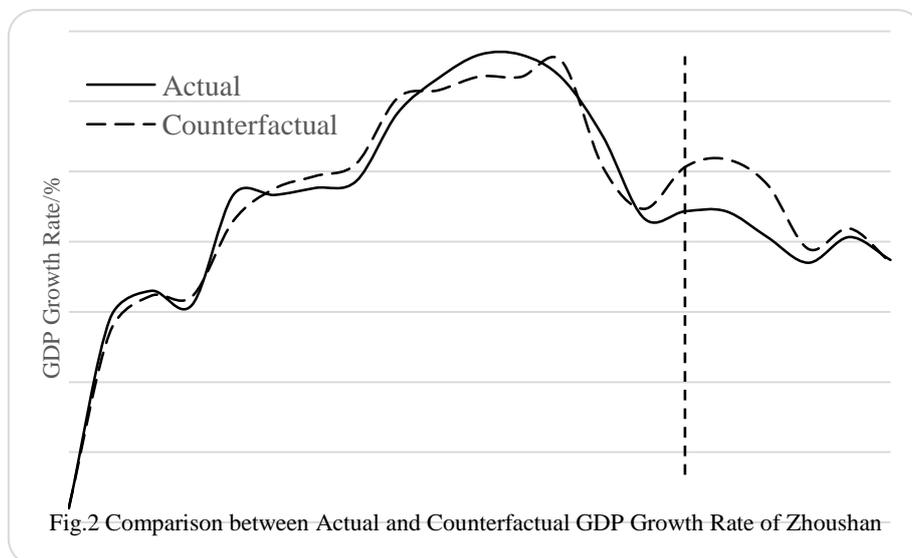

Fig.2 Comparison between Actual and Counterfactual GDP Growth Rate of Zhoushan

Based on the above analysis, we computed Zhoushan's counterfactual GDP growth rate from 2011 to 2015 by formula (7). All the counterfactual and actual values were listed in Figure 2 (on the right side of the vertical line in Figure 2, i.e. after 2010). By subtracting counterfactual values from actual values, the treatment effect of ZANA is estimated and listed in Table 2. We find the treatment effects are negative and their absolute values gradually decreased from 2011 to 2014. Until 2015 the treatment effect is turned into a positive. But its mean in the five years still is negative, -1.07%. It shows that ZANA had some negative effect on GDP growth rate at its beginning, then the negative effect gradually diminished and disappeared in five years later. So we have reason to believe that the policy of ZANA does not improve the quantity of Zhoushan's economic growth. Next we want to know how does the policy impact on the quality of Zhoushan's economic growth.

Tab.2 Treatment effect of ZANA for GDP growth rate in Zhoushan

| year | Real growth rate（%） | counterfactual growth rate（%） | treatment effect（%） |
|---|---|---|---|
| 2011 | 11.3 | 13.53 | -2.22923 |
| 2012 | 10.2 | 12.45 | -2.25266 |
| 2013 | 9.1 | 9.69 | -0.5946 |
| 2014 | 10.2 | 10.57 | -0.36948 |
| 2015 | 9.22 | 9.11 | 0.10628 |
| mean | 10.00 | 11.07 | -1.07 |

**4.2.2** Empirical analysis on economic efficiency

In order to analyze the policy impact of ZANA on local economic quality, the economy efficiency is given as an output index. With sup-efficiency of 20 cities from 1995 to 2010, which has been calculated in 3.2 and listed in Tab.1, the best fitting of Zhoushan's economic efficiency (y) before ZANA is chosen by the above three steps and expressed as follows:

$$y=-0.9885+0.5487x_1+0.4392x_2+0.5269x_3+0.3232x_4-0.3792x_5+0.2396x_6 \qquad (8)$$

Where $x_1$ to $x_6$ represent the economic efficiency of Nanjing ($x_1$), Zhenjiang ($x_2$), Wenzhou ($x_3$), Bengbu ($x_4$), Huaibei ($x_5$) and Chuzhou ($x_6$) respectively. The six cities are composed of a control group. The regression coefficient $R^2$ to formula (8) is 0.998. Its equation's overall significance test (F test) and the variable's significance test (t test) are also passed well. The fitted and the actual economic efficiency are shown in Figure 3 for comparing (in the left side of the vertical line in Figure 3, i.e. before 2010). It can be seen that the fitted values derived from the control group can fit Zhoushan's actual values well between 1995 and 2010.

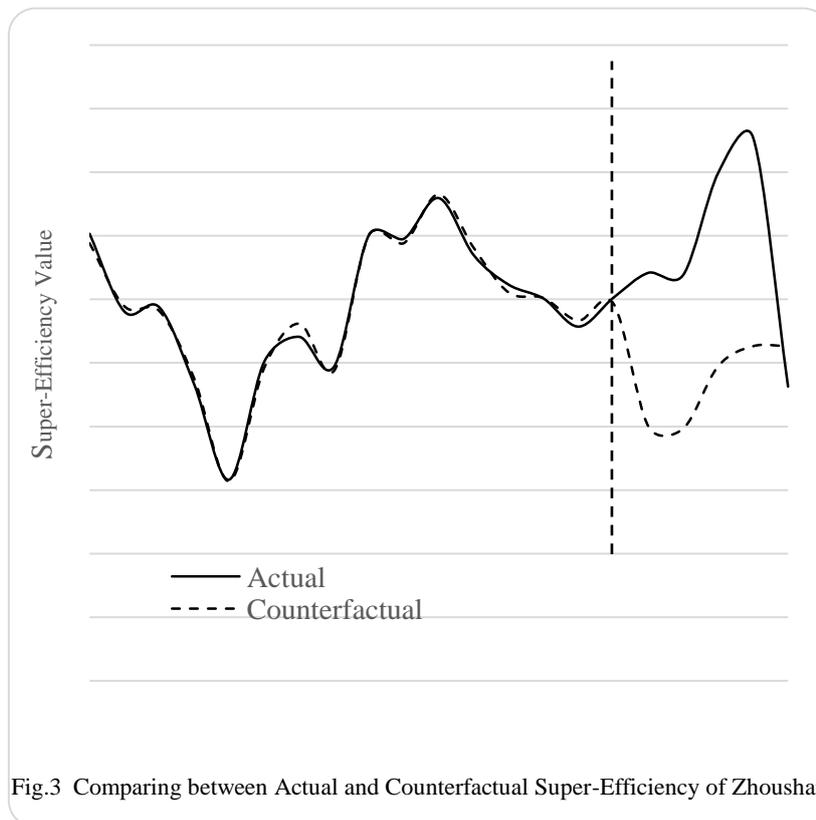

Fig.3 Comparing between Actual and Counterfactual Super-Efficiency of Zhoushan

By formula (8), the out-of-sample for Zhoushan, which is counterfactual super-efficiency values without interference of ZANA in 2011-2015, could be predicted. The counterfactualand actual paths are shownseparatelyinFigure 3 (on the right side of the vertical line inFigure 3, i.e. after 2010).The treated effect ofZANAcould be obtainedby subtracting counterfactual values from the actualvalues. The results are positive from 2011 to 2014, and has increased from 0.2226 to 0.3049in Table 3. It shows that the economic efficiency of Zhoushan has been improved significantly in these years.ZANAhas indeed effectively promoted the economic development quality of Zhoushan. But in 2015 the treated effect almost got close to 0.Itshows that the positive treated effect ofZANA disappeared after four years.

Comparing table 3 with table 2, we get a broad view of the impact of ZANA on Zhoushan's economy. ZANA has negatively affected local GDP growth rate, but improvers local economic efficiency. After four years, the effect of ZANA on GDP growth rate and economic efficiency is all weakened and disappeared gradually. It is said that the direct impact period of ZANA on Zhoushan's economy isn't long.

### 4.3 Robustness check of the used model

We noted that there are some assumptions about Hsiao's panel data approach. According to this approach, much of the cities' economy in the control group must be driven by the same common factors as the ones of Zhoushan. So the robustness of the approach in this paper is that a fundamental relation among all the studied cities must remain unchanged from pre intervention to post intervention and should not be related to a specific time point. To verify the treatment effects estimated in Sections 3.2 is not related to the time point of 2010, we select any time point before ZANA replace the year of 2010 and then take same counterfactual analysis. If the fitted results are still as same as the above analysis, the approach will be proven stable in this paper and its results will be reliable.

**Tab. 3 Treated effect of ZANA fo rsuper efficiency in Zhoushan**

| Year | Real efficiency | Counter-factual efficiency | Treatment effect |
|---|---|---|---|
| 2011 | 0.6415 | 0.4003 | 0.2412 |
| 2012 | 0.6383 | 0.3969 | 0.2414 |
| 2013 | 0.7987 | 0.4949 | 0.3038 |
| 2014 | 0.8542 | 0.5263 | 0.32795 |
| 2015 | 0.4627 | 0.5264 | -0.0637 |
| mean | 0.67908 | 0.4690 | 0.2101 |

We verified the approach from both sides of the GDP growth rate and economic efficiency. Without losing generality, we assume the approved time point of ZANA two years ahead of schedule. So in the subsequent analysis, we fit Zhoushan's economy from 1995 to 2008 and do its counterfactual analysis from 2009 to 2015.

**Tab.4 Optimal fitting coefficient and its test value of GDP Growth Ratefrom 1995 to 2008**

| Control Group | Coefficient | Std. Error | t-value | t (p) |
|---|---|---|---|---|
| Changzhou ($X_1$) | 1.9535 | 0.1147 | 17.0288 | 0 |
| Nantong ($X_2$) | -0.5601 | 0.0407 | -13.7772 | 0 |
| Wenzhou ($X_3$) | -0.5330 | 0.0643 | -8.2833 | 0 |

Firstly for GDP growth rate, similar to the section 3.2.1, in the given 19 cities, Changzhou ($X_1$), Nantong ($X_2$) and Wenzhou ($X_3$) are selected to form a control group by Pampe Package in R software , Their regression equation for Zhoushan (y) was gotton as follows:

$$y=3.201752+1.953512X_1-0.560051X_2-0.532979X_3 \quad \quad \quad (9)$$

The test values of formula (9) are listed in table 4. They are a ideal result. By formula (9), the counterfactual GDP growth rate without the interference of ZANA has been predicted from the year of 2009 to 2015. All the fitted and actual values are shown in Figure 4. They are fittedin well between 1995 and 2008 (in theleft side of the vertical line in Figure 4, i.e. before 2008). Furthely, comparing the counterfactual values in Figure 2 and Figure 4 from 2010 to 2015, both of them have similar values and trends, that is, the counterfactual values are larger than and gradually close to the actual values. This shows that the approach is stable and reliable for the analysis of GDP growth rate in this paper.

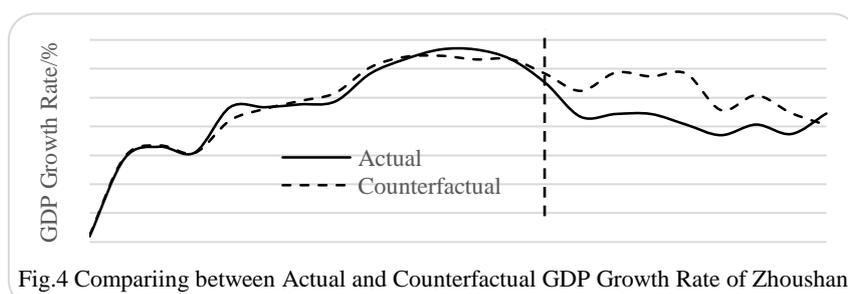

Fig.4 Compariing between Actual and Counterfactual GDP Growth Rate of Zhoushan

Secondly, just as the counterfactual analysis in section 3.2.2, Nanjing ($X_1$), Zhenjiang ($X_2$), Wenzhou ($X_3$), Bengbu ($X_4$), Huaibei ($X_5$), Chuzhou ($X_6$) are selected in the given 19 cities by Pampe Package in R softwareto form a control group, Their regression equation for economic efficiency of Zhoushan (y) is gotten as follows:

$$y=-0.9865+0.5312X_1+0.4433X_2+0.5294X_3+0.3254X_4-0.3748X_5+0.2383X_6 \quad \quad (10)$$

Tab.5 Optimal fitting coefficient and its test value of economic efficiency for Zhoushan from 1995 to 2008

| Control Group | Coefficient | Std.Error | t-value | t (p) |
| --- | --- | --- | --- | --- |
| Nanjing ($X_1$) | 0.5312 | 0.1107 | 4.800 | 0.002 |
| Zhenjiang ($X_2$) | 0.4433 | 0.0475 | 9.335 | 0 |
| Wenzhou ($X_3$) | 0.5294 | 0.0462 | 11.4489 | 0 |
| Bengbu ($X_4$) | 0.3254 | 0.0313 | 10.3914 | 0 |
| Huaibei ($X_5$) | -0.3748 | 0.0552 | -6.7904 | 0 |
| Chuzhou ($X_6$) | 0.2383 | 0.0227 | 10.5014 | 0 |

The test values of formula (10) are listed in table 5. By formula (10), counterfactual economic efficiency without the interference of ZANA is predicted from 2009 to 2015. Then all the fitted and actual values are shown in Figure 5. They are shown to fit well between 1995 and 2008 (in the left side of the vertical line in Figure 5, i.e. before 2008). Comparing the counterfactual values in Figure 3 and Figure 5 from 2010 to 2015, we find that both of them have similar values and trends, that is, the actual values were larger than the counterfactual values between 2011 and 2014, and in 2015 they were the same. This shows that the approach in this paper is stable and reliable for the analysis of economic efficiency too.

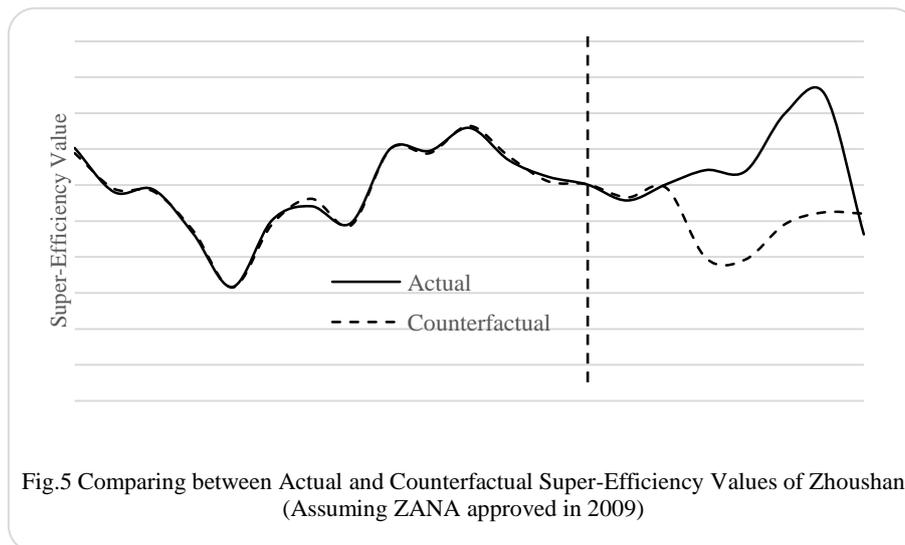

Fig.5 Comparing between Actual and Counterfactual Super-Efficiency Values of Zhoushan (Assuming ZANA approved in 2009)

## 5  Conclusions and discussion

In this paper, the input and output data of 20 cities in Jiangsu, Zhejiang and Anhui provinces of China is collected from 1995 to 2010. The economic efficiency of these cities is measured by the super-efficiency DEA model with these data. Furthermore, the GDP growth rate and economic efficiency of Zhoushan are analyzed by Hsiao's panel data approach to study the impact of ZANA,. The research shows that: in the first four years, ZANA promoted Zhoushan's economic quality by raising its economic efficiency, but relatively reduced its economic quantity by negatively affecting its GDP growth rate. However, the impact on Zhoushan's GDP growth rate and economic efficiency gradually disappeared in the fifth year.

It is somewhat unexpected that the policy of NNA doesn't promote local GDP growth rate. Is this a special case?

We found similar results in other studies, although those authors did not pay special attention to their causes. Fan et. al. (2018) found that some NNAs in south China, such as Liangjiang New Area, Nansha New Area, Guian New Area, Tianfu New Area and Xiangjiang New Area, had negative influence on GDP growth rate of their hosting provinces. Chao et al. (2018) also found that the contribution to GDP growth of NNAs in Chinese eastern regions is significantly less than that in the central and western regions. These studies simply attributed these phenomena to the geographical location of the NNAs. In addition, some studies have also shown that the impact of NNA on GDP growth is greater in the past than now. Analyzing these research and combining our study on Zhoushan in this paper, we find that the Geographical location or establishment time of NNAs is only a superficial reason, the nature of these phenomena is due more to the economic development level of these NNAs' hosting areas. The policy of NNAs has less impact for GDP growth if they are approved in the developed region rather than in undeveloped region. It is said that approving a NNA in relatively underdeveloped areas will more improve local economic growth rate.

This study also shows that ZANA improves local economic efficiency in a certain period. However, there are few case studied in this area, so whether this conclusion is right for other NNAs needs to be further empirical research. But theoretically, we think it is reasonable. Because the policy advantages of NNAs gathered resources and enterprises (Wang et al., 2019), the concentration improves local economic efficiency. But Wang et al (2016) found that the "agglomeration effect" lasts for a very short time and gradually disappears about three years after the establishment of the NNA. The case of Zhoushan in this paper basically verified this conclusion.

Based on the above analysis, we think that it is better to set up a NNA in a relatively undeveloped zone. We suggest that when a NNA is approved, the local government should make full use of the policy benefits to adjust its industrial structure and promote advantageous industries as soon as possible, otherwise the effective time of the policy will not be too long.